\documentclass{llncs}

\usepackage[english]{babel}
\usepackage{blindtext}
\usepackage{xcolor}
\usepackage[single]{acro}
\usepackage{booktabs, caption, tabularx, makecell}
\usepackage{multirow}
\usepackage{tablefootnote }

\DeclareAcronym{soc}{short = SoC, long = System-on-a-Chip, foreign-lang = english,	single-format = \MakeLowercase}

\DeclareAcronym{ecc}{
	short = ECC,
	long  = Error-Correcting Code,
	single-format = \MakeLowercase
}

\DeclareAcronym{iot}{
	short = IoT,
	long  = Internet of Things,
	single-format = \MakeLowercase
}

\DeclareAcronym{cba}{
	short = CBA,
	long  = Circuit Board Assembly,
	single-format = \MakeLowercase
}

\DeclareAcronym{ic}{
	short = IC,
	long  = integrated circuit,
		long-plural-form  = integrated circuits,
	single-format = \MakeLowercase
}

\DeclareAcronym{oem}{
	short = OEM,
	long  = Original Equipment Manufacturer,
	single-format = \MakeLowercase
}

\DeclareAcronym{gpu}{
	short = GPU,
	long  = Graphics Processing Unit,
	long-plural-form  = Graphics Processing Units,
	single-format = \MakeLowercase
}

\DeclareAcronym{ea}{
	short = E/A,
	long  = Eingabe/Ausgabe ,
}
\DeclareAcronym{cots}{
	short = COTS,
	long  = Commercial off-the-shelf,
	single-format = \MakeLowercase
}
\DeclareAcronym{chwc}{
	short = CHWC,
	long = \ac{cots} Hardware Component ,
	foreign-lang = english,
	single-format = \MakeLowercase
}

\DeclareAcronym{chwca}{
	short = CHWCA,
	long = COTS Hardware Component Assurance,
	foreign-lang = english,
	single-format = \MakeLowercase
}

\DeclareAcronym{da}{
	short = DA,
	long = Design Assurance,
	long-plural-form  = Design Assurance,
	foreign-lang = english,
	single-format = \MakeLowercase
}

\DeclareAcronym{dan}{
	short = DAN,
	long = Development Assurance Niveau,
	long-plural-form  = Development Assurance Niveau,
	foreign-lang = english,
	single-format = \MakeLowercase
}

\DeclareAcronym{seh}{
	short = SEH,
	long = Simple Electronic Hardware,
	long-plural-form  = Simple Electronic Hardware,
	foreign-lang = english,
	single-format = \MakeLowercase
}

\DeclareAcronym{ceh}{
	short = CEH,
	long = Complex Electronic Hardware,
	long-plural-form  = Complex Electronic Hardware,
	foreign-lang = english,
	single-format = \MakeLowercase
}

\DeclareAcronym{smcu}{
	short = SMCU,
long = Safety Microcontroller Unit,
long-plural-form  = Safety Microcontroller Units,
foreign-lang = english,
single-format = \MakeLowercase
}

\DeclareAcronym{cri}{
	short = CRI,
	long  = Certification Review Item ,
}

\DeclareAcronym{ip}{
	short = IP,
	long  = Intellectual Property,
	long-plural-form  = Intellectual Properties,
}

\DeclareAcronym{ee}{
	short = e/e,
	short-plural-form = e/e,
	long  = elektrische/elektronische,
	long-plural-form = elektrischen/elektronischen,
	single-format = \MakeLowercase
}

\DeclareAcronym{uav}{
	short = UAV,
	long  = Unmanned Air Vehicle,
	single-format = \MakeLowercase
}

\DeclareAcronym{amc}{
	short = AMC,
	long  = Acceptable Means of Compliance,
	single-format = \MakeLowercase
}
\DeclareAcronym{aeh}{
	short = AEH,
	long  = Airborne Electronic Hardware,
	long-plural-form = Airborne Electronic Hardware,
	short-plural-form = AEH,
	single-format = \MakeLowercase
}
\DeclareAcronym{fc}{
	short = FC,
	long  = failure condition
}
\DeclareAcronym{easa}{
	short = EASA,
	long  = European Aviation Safety Agency,
	single-format = \MakeLowercase
}
\DeclareAcronym{cs}{
	short = CS,
	long  = Certification Specification,
		single-format = \MakeLowercase
}

\DeclareAcronym{mcu}{
	short = MCU,
	long = Microcontroller Unit,
	foreign-lang = english,
		single-format = \MakeLowercase
}

\DeclareAcronym{asic}{
	short = ASIC,
	long = anwendungsspezifische integrierte Schaltung,
	foreign  = Application-Specific Integrated Circuit,
	long-plural = en,
	foreign-lang = english
}

\DeclareAcronym{amcu}{
	short = AMCU,
	long = Automotive Microcontroller Unit,
	foreign-lang = english,
		single-format = \MakeLowercase
}

\DeclareAcronym{mpu}{
	short = MPU,
	long = Microprocessor Unit,
		single-format = \MakeLowercase
}

\DeclareAcronym{wcet}{
	short = WCET,
	long  = Worst Case Execution Time,
	single-format = \MakeLowercase
}

\DeclareAcronym{ccdl}{
	short = CCDL,
	long  = Cross Channel Data Link,
	single-format = \MakeLowercase
}

\DeclareAcronym{dmac}{
	short = DMA-C,
	long  = Direct Memory Access Controller,
	single-format = \MakeLowercase
}

\DeclareAcronym{mmu}{
	short = MMU,
	long  = Memory Management Unit,
	single-format = \MakeLowercase}
\DeclareAcronym{ima}{
	short = IMA,
	long  = Integrated Modular Avionic,
	single-format = \MakeLowercase
}

\DeclareAcronym{ecmp}{
  short = ECMP,
  long  = Electronic Component Management Plan,
    single-format = \MakeLowercase
}

\DeclareAcronym{seooc}{
	short = SEooC,
	long  = Safety Element out of Context,
	single-format = \MakeLowercase
}
\DeclareAcronym{swap}{
	short = SWaP,
	long  = {Space, Weight and Power},
	single-format = \MakeLowercase
}

\DeclareAcronym{hwdal}{
	short = HWDAL,
	long  = Hardware Design Assurance Level,
	single-format = \MakeLowercase
}

\DeclareAcronym{dal}{
	short = DAL,
	long  = Development Assurance Level,
	single-format = \MakeLowercase
}

\DeclareAcronym{pse}{
	short = PSE,
	long  = Product Service Experience,
	single-format = \MakeLowercase
}

\DeclareAcronym{lru}{
	short = LRU,
	long  = Line Replaceable Unit,
	single-format = \MakeLowercase
}

\DeclareAcronym{asil}{
	short = ASIL,
	long  = Automotive Safety Integrity Level,
	single-format = \MakeLowercase
}

\DeclareAcronym{fcc}{
	short = FCC,
	long  = Flight Control Computer,
	single-format = \MakeLowercase
}

\DeclareAcronym{bist}{
	short = BIST,
	long  = Build In Self Test,
	single-format = \MakeLowercase
}

\DeclareAcronym{aadl}{
	short = AADL,
	long  = Architecture Analysis and Design Language
}

\DeclareAcronym{faa}{
	short = FAA,
	long  = Federal Aviation Administration,
	single-format = \MakeLowercase
}

\DeclareAcronym{seu}{
	short = SEU,
	long  = Single Event Upset,
	single-format = \MakeLowercase
}

\DeclareAcronym{ca}{
	short = CA,
	long  = Certification Authority,
	single-format = \MakeLowercase
}
\begin{document}
\title{Assurance Benefits of ISO 26262 compliant Microcontrollers for safety-critical Avionics}
\author{Andreas Schwierz \inst{1} \and H\r{a}kan Forsberg \inst{2}}

\institute{Research Center:\\Competence Field Aviation\\ Technische Hochschule Ingolstadt\\ 85049 Ingolstadt, Germany\\ \email{ Andreas.Schwierz@thi.de}
	\and
School of Innovation, Design and Engineering\\Division of Intelligent Future Technologies\\ M\"alardalen University\\ 721 23 V\"aster\r{a}s, Sweden\\ \email{hakan.forsberg@mdh.se}}

\maketitle            

\begin{abstract}
The usage of complex \acp{mcu} in avionic systems constitutes a challenge in assuring their safety. They are not developed according to the development requirements accepted by the aerospace industry. These \ac{cots} hardware components usually target other domains like the telecommunication branch. In the last years \acp{mcu} developed in compliance to the ISO 26262 have been released on the market for safety-related automotive applications. The avionic assurance process could profit from these safety \acp{mcu}. In this paper we present evaluation results based on the current assurance practice that demonstrates expected assurance activities benefit from ISO 26262 compliant \acp{mcu}.
\end{abstract}

\section{Introduction}
\ac{cots} hardware components are ubiquitous in \ac{aeh} and were considered in the very beginning of the RTCA/DO-254~\cite{RTCA.19.4.20001}. However, the complexity of the desired COTS components is continuously increasing, even for highly safety-critical functions. Certification authorities address this rapid evolution by delegation of research activities and provision of further guidance in \ac{cots} hardware component assurance for different types of components (e.g. \acp{mcu} or \acp{gpu}). The aim is to deliver advisory material as specific as possible for industrial practice. \ac{cots} hardware component assurance and \ac{da} of \ac{aeh} have the same objective, which is to assure that a hardware component safely performs as intended in its operational context. But the method is inevitably distinct because of the nature that \ac{cots} hardware components were not developed according to the RTCA/DO-254 or that \ac{cots} manufacturers do not disclose required development artefacts to be able to demonstrate compliance afterwards. So process-based evidence of the design life cycle cannot be claimed as aircraft systems concerns were not regarded during the development of the \ac{cots} product. 

Avionic manufacturers employ components actually intended for other domains\footnote{For \ac{dal} A applications the general principle is to restrict the use of complex components.}. Hardware with a long market availability and operable under harsh environmental conditions is requested. These component properties are characteristic for the automotive domain. Functional safety is at least since 2011, where the ISO 26262 standard~\cite{ISO.2011} has been released, a major concern for \acp{oem} and also many suppliers for automotive parts like \acp{ic}. \acp{mcu} developed in compliance with ISO 26262 are designed for safety-critical applications. For their development the ISO 26262 describes an approach called \ac{seooc}. Semiconductor manufacturers are able to create a product that can be integrated into different systems or operational contexts. \ac{aeh} manufacturers observe the situation in the automotive and other safety-critical domains that request hardware components according to standards that are aimed to reduce or control the risk of hazardous failures~\cite{Schwierz.2017b}.Their aim is to exploit the fact that safety plays an essential role in more and more sectors and to influence the product lines of hardware component manufacturers that produce in high quantities. 

As a consequence of the current situation, the following research question arises: How can the avionics industry benefit from this situation in the course of \ac{cots} hardware components assurance? To answer this question, the paper is structured as following: Section \ref{sec:assuranceInAv} describes how assurance is achieved for avionic systems in general and how it differs if complex \acp{mcu} shall be embedded. Evaluation of ISO 26262 compliant \ac{mcu} benefit in \ac{cots} hardware component assurance is performed in section \ref{sec:isoreuse}. Conclusions are given in the last section.

\section{Assurance Methods for Avionics}\label{sec:assuranceInAv}
The meaning of the concept of assurance varies in its understanding depending on the context and which aspects should be assured. The following sections give a brief overview of this topic in the avionics domain and distinguish between two aspects. First, these are avionic systems that mainly comprise components manufactured alongside the avionic development life cycle and second avionic systems that make use of complex \acp{mcu}. In both cases sufficient and practical assurance methods have to be performed.  

\subsection{Development Assurance}
The term assurance methods is currently often used in the avionics domain~\cite{Mutuel.2017,DeWalt.2014,Jean.}. In general, assurance can be defined as the actions that provide appropriate confidence and evidence that a product or process meets its requirements~\cite{SAEAerospace.12.2010}. Assurance intends to reduce the uncertainty about the correct realisation of the product.
It delivers reasons why the confidence on achieving the claim is so justifiable~\cite{ISO.2013} and why most assurance activities target the establishment of this confidence~\cite{Holloway.2015}. In a requirements-based product development this means, that the requirements specification meets the real-world needs (\emph{validation of requirements}) and that the product is a correct implementation of the requirements specification (\emph{verification of requirements}).

For avionic systems the airworthiness requirements~\cite{EASA.2016} are on the top of their requirements specification. Summarized, it has to be assured that the avionic system design is appropriate for the intended function and that its function is provided as defined in its operational context (environmental and operating conditions of the aeroplane). These are prerequisites to ensure that it is extremely improbable that safety-critical \ac{aeh} contributes to a catastrophic failure condition at aircraft level that harms human life. For safety-critical systems, assurance methods are necessary to deliver enough credit to justifiably state that the system is safe in its context\footnote{For non-safety-critical systems other properties e.g. security are in the focus of assurance methods.}. 

This is very challenging if the system is too complex in order to provide the requested level of confidence by exhaustive tests which fully characterise the system. Hence, the method of \emph{\acl{da}} was defined to cope with this issue in different areas, system~\cite{SAEAerospace.1996,SAEAerospace.12.2010}\footnote{The safety assessment is part of development assurance to deliver the safety requirements and support the confidence of its verification.}, software~\cite{RTCA.03.131} and hardware~\cite{RTCA.19.4.20001} development. These \ac{da} areas aim to accomplish the development in a sufficiently rigorous and disciplined way so that development errors do not impact safety~\cite{AIR120.2006}. \ac{da} is characterised by techniques that are applied during the whole development process in order to identify and correct errors that could occur at various steps within the development life cycle. This comprises assurance techniques like process assurance, verification coverage criteria and reviews. In addition, for the two highest development assurance levels, RTCA/DO-254 requires additional development assurance activities to be performed, i.e. it is not sufficient to show evidence that a certain design process has been followed alone. RTCA/DO-254 suggests using architectural mitigation techniques, service experience and advanced verification methods as additional development assurance activities.

For \ac{aeh} embedded \acp{mcu}, \ac{da} on device level cannot be claimed or used as assurance method~\cite{AVSI.06.2008}. The reason is that most \ac{da} techniques are based on an ongoing development and the accessibility of development-time artefacts down to a level detailed enough to realize the hardware component and assure its safety aspects. For \acp{mcu} this is not possible as the development has already been accomplished and detailed development-time artefacts are not available. Thus, other assurance methods have to be determined which can reduce uncertainty in a similar magnitude creditable by the certification authorities.

\subsection{COTS Hardware Component Assurance}\label{sec:cotsAssurance}
As stated in~\cite{AFE75ProjectManagementCommittee.}, it is very challenging to define an assurance method for \ac{cots} hardware components in an objective way. Two aspects have to be considered for the objectivity: The method is applicable for a variety of components and following it delivers results that can be fairly assessed by the authorities. Such guidance shall support the industry in realization of certifiable \acp{aeh}, embedded with \ac{cots} components, in a practical way so that certification costs do not explode and safety can be sufficiently assured. 

The latest initiatives by authorities in this direction resulted in the following documents\footnote{None of these listed documents are binding guidance material.} and reflect the status quo:
\acuse{easa,faa}
\begin{itemize}
	\item \emph{\acs{easa}:} 
	\begin{itemize}
		\item Certification memorandum CM SWCEH-001: Development Assurance of Airborne Electronic Hardware~\cite{EASA.9.5.20121}.
	\end{itemize}
This document represents the current attitude of the \ac{easa} about several certification aspects of \ac{aeh} and in section 9 especially to \ac{cots} \acp{mcu}. The content is based on experience in \ac{cots} hardware component assurance in many certification projects gathered in the years before and funded research activities as~\cite{EASA.15.2.2008}.
	\item \emph{\acs{faa}:}
	\begin{itemize}
		\item Commercial Off-The-Shelf Airborne Electronic Hardware Assurance Methods - Phase 3 - Embedded Controllers~\cite{Mutuel.2017}.
		\item Assurance of Multicore Processors in Airborne Systems~\cite{Mutuel.2017b}.
	\end{itemize}
These technical reports are the results of funded research by the \ac{faa} to develop proposals for assurance approaches for different \ac{cots} hardware.
\end{itemize}

Notable is that for different hardware component categories (e.g. multicore processors or microcontrollers) the assurance methods were separately considered. This faces the fact, that each technology has its own issues which shall be incorporated to provide methods that are useful in practice. Especially small companies and market newcomers are interested in guidelines as concrete as possible since they do not have the same amount of development experience as the larger ones~\cite{Strasburger.2014}.

All reports listed above share one similarity: \ac{cots} assurance should be managed from system level in parallel or within the \ac{aeh} design process\footnote{Recommended also by RTCA/DO-254.}. However, they lack in formulating a framework that brings them all together in a coherent approach that could be related as an deployable \ac{cots} hardware component assurance process. From these reports, the \ac{easa} certification memorandum can be considered as most relevant to identify necessary \ac{cots} assurance tasks. It represents the current position of a certification authority and defines assurance activities as an \ac{ecmp}\footnote{The term \ac{ecmp} in the memorandum is misleading because typically such a process does not perform profound functional assurance activities.} extension.

FAA’s research report~\cite{Mutuel.2017} does not explicitly cover activities but has identified issues to keep track of during assurance. Also, findings and recommendations can be found in this research report. Some similar to the activities in EASA’s CM, e.g. usage domain analysis, integration aspects, errata handling, and configuration management, and some similar to typical ISO 26262 implementations, e.g. robustness verification. These issues, findings and recommendations have been analysed but were not identified as obvious COTS assurance tasks and therefore not included in this paper. In addition, the other FAA research report~\cite{Mutuel.2017b} concerns multicore processors running in parallel and not in synchronized lock-step mode (LSM). For the case study in this paper, all ISO 26262 developed \acp{mcu} must be run in LSM for safety-critical applications. Report~\cite{Mutuel.2017b} is therefore not included in the analysis of this paper.

The memorandum contains sixteen recommended activities, numbered with brackets from 1 to 16 (e.g. Activity [1]). These activities are referenced in this article with the same numbering but are emphasized in round brackets instead to avoid confusion with other references in the article. These activities should be considered depending on the \ac{dal} associated by a higher level safety assessment, the magnitude of \ac{pse} traceable from different domains and the complexity of the \ac{mcu}. In the subsequent text, activities are discussed for \ac{dal} A which apply for components with the highest possible safety impact. This will extend the value of scope of this article, because targeting \ac{dal} A means that all activities have to be conducted if the \ac{pse} is inadequate. The argumentation behind these additional assurance activities is not further stated in the document but is essential for the understanding on how they contribute to \ac{cots} hardware component assurance. Thus the assumed argumentation was reconstructed. 

Two top arguments were extracted that have to be assured:
\begin{description}
	\item[Argument 1] The component performs as described by the manufacturer without anomalous behaviour. 
	\item [Argument 2] The component as used satisfies the \ac{aeh} requirements.
\end{description}
It has to be differentiated between those arguments as the \ac{mcu} was not developed according to the requirements of the \ac{aeh}. How these arguments can be supported depends on the complexity of the \ac{cots} component. For \acp{mcu} with a functional architecture classified as simple, the arguments can be fulfilled as following:

\begin{itemize}
	\item \emph{Argument 1:}
	\begin{itemize}
		\item Verification of component behaviour on device level as specified by the manufacturer.\vspace{4pt}
		
		The simplicity of the \ac{cots} component allows to verify  all requirements on the physical device.
		\item Substantiate the confidence of a design free from anomalous behaviour by demonstrating device maturity or quality. \vspace{4pt}
		
		Most of the confidence on device quality is already supported by the comprehensive verification effort. However, additional errata management in activity (6) and (7) shall be considered to state that the device design is stable enough. This can be demonstrated by errata decreasing over the service time on the market. Also the errata publishing policy of the manufacturer shall be adequate to be always informed about revealed problems and to achieve that errata with potential safety impacts can be handled. 
	\end{itemize}
\item \emph{Argument 2:}
\begin{itemize}
	\item Verification of \ac{aeh} requirements on \ac{lru} or \ac{cba} level during equipment design.
	\item As requested by certification requirements, no single point of failure should lead to a catastrophic failure condition. This is also valid for \ac{cots} components in general. Activity (15) requests the implementation of an adequate architectural mitigation technique like dissimilar redundancy or monitoring.
	\item An \ac{ecmp} e.g. as described in IEC TS62239.
\end{itemize}
\end{itemize} 

Most of the available \acp{mcu} on the market are complex or even highly complex components. For these devices, exhaustive tests on device level can not be achieved to adequately substantiate argument 1 as for simple components. Therefore additional activities for complex or highly complex hardware are necessary, which are depicted as following for argument 1 and 2:
\begin{itemize}
	\item \emph{Argument 1:}
	\begin{itemize}
		\item Verification of component behaviour on device level as specified by the manufacturer.\vspace{4pt}
		
		The concept of usage domain as described in activity (4) resp. (5) is suggested to bound the scope of device level verification only on component behaviour that is relied on or is really used. The determined usage domain shall be compliant to the manufacturer recommendations and verified on device level. If the \ac{mcu} is part of a partitioning concept, an analysis has to be performed as described in activity (16) to claim the robustness of this mechanism at device level\footnote{Actually, we consider partitioning aspects as a specific part of the usage domain analysis, because \ac{mcu} properties shall be verified on device level.}.
	
		\item Substantiate the confidence of a design free from anomalous behaviour by demonstrating device maturity or quality. \vspace{4pt}
		
		The verification on device level limited to usage domain aspects is not enough to mainly support argument 1. In comparison to simple \ac{cots} components, the correct behaviour assumption of complex hardware is more based on other activities like:
		\begin{itemize}
			\item \ac{cots} manufacturer quality management and production process has to be assessed in activity (3).
			\item Errata management as for simple components in activity (6) and (7). Additionally, activity (8) requests that the \ac{aeh} manufacturer has to document own made experience with the hardware during the development (e.g. errata workarounds).
			\item Manufacturers configuration management including a change process has to be assessed in activity (9) to make sure that changes are appropriately controlled and communicated. Activity (10) additionally requests a change impact analysis to identify potential extra verification effort.
			\item The \ac{pse} has to be documented by activity (13) in order to determine if it is sufficient\footnote{The metric to determine a \ac{pse} as sufficient is also defined in the certification memorandum.}to omit certain assurance activities. Specifically for \ac{dal} A and B, a minimum amount of \ac{pse} has to be reported in order to exclude really novel designs to be embedded in \ac{aeh} systems. Activity (14) further increases the confidence on the maturity and stability of the \ac{mcu} by requesting evidence on the rate and fact of past modifications.			
		\end{itemize}
	\end{itemize}
\item \emph{Argument 2:} 
\begin{itemize}
	\item Usage domain validation in activity (5) ensures that the usage domain is consistent to system, software and hardware requirements. 
	\item For complex \ac{cots} it is not sufficient to verify requirements allocated to the \ac{mcu} at equipment level as for simple components. Activity (11) requests verification and validation of these requirements coming from other hardware or software components on device level in order to get confidence about its correct integration.
	\item For highly complex \acp{mcu} activity (12) has to be conducted to have a clear understanding of possible device failure modes and rates depending on its configuration. 
	\item Architectural mitigation technique as requested by activity (15) shall also be applied.
		\item An \ac{ecmp} e.g. as described in IEC TS 62239~\cite{IEC.2015}.
\end{itemize}
Activity (1) and (2) were not mapped to a top argument, since determining or classifying the \ac{mcu} characteristic (1) and archiving public available device data (2) are required for both top arguments. It does not matter if the \ac{mcu} is classified as simple, complex or highly complex.
\end{itemize}
 All these explained assurance activities of the certification memorandum are only applicable for the peripheral subsystem and other functions which are not part of the processing core. The \ac{da} of the processing core is based on the software development process compliant to RTCA/DO-178 that includes software testing on the target hardware platform. This separation is based on the assumption that other \ac{mcu} functions do not interfere with the software execution on the processing core~\cite{Mutuel.2017}.

The explanations about complex \ac{cots} hardware component assurance established the basis on which in the next section the potential benefits from ISO 26262 compliant complex \acp{mcu} can be examined.

\section{Benefits from ISO 26262 compliant MCUs in AEH COTS Assurance}\label{sec:isoreuse}
The research question asked in the introduction was: \emph{How can the avionics industry benefit from ISO 26262 compliant \acp{mcu} in the course of \ac{cots} hardware components assurance?} Before starting to evaluate an ISO 26262 compliant \ac{mcu} against the assurance approach from section \ref{sec:cotsAssurance}, the differences to other \acp{mcu} on the market have to be identified first. What makes these \acp{mcu} so special? These are the aspects on which \ac{cots} assurance can probably profit in comparison to other \acp{mcu} e.g. from the telecommunication domain. 
\subsection{Determination of ISO 26262 specifics for reuse}\label{sec:isoSpec}
The special characteristics of interest come from the development approach defined by the process requirements from ISO 26262. During previous research we made a comparison between the \ac{da} method of RTCA/DO-254 and ISO 26262-5\footnote{Part five of the standard is about product development at the hardware level.}, which concludes that the ISO 26262 does not reach the same level of design integrity~\cite{Schwierz.2017}\footnote{This demonstrates reasonableness of a dedicated \ac{cots} assurance process see section \ref{sec:cotsAssurance}.}. The reason is that only safety requirements are considered in the development life cycle of the \ac{mcu}, whereas the traceability down to detailed design level is not required. For manufacturers the main focus is on the safety architecture to handle random hardware failures by adequate safety mechanisms to achieve the targeted diagnostic coverage and to be able to enter a safe state if necessary or indicate failures to external components. Thus, the main focus is not on systematic errors\footnote{ISO 26262 implicitly addresses systematic errors for hardware through the development process.}, which is the main focus for designs following RTCA/DO-254. On the device level the characteristic of a very high diagnostic coverage makes these products something special on the market and manufacturers are very encouraged in the realization and verification of the \ac{mcu}'s safety architecture. 

The \ac{mcu} development approach has to adhere to ISO 26262-5 and referenced parts. ISO 26262-10:2012 does not define conformance requirements but gives guidance especially on \ac{mcu} development. It explains the \ac{seooc} method and describes in appendix A how it could be applied for \acp{mcu}. This concept allows the realization of a component like an \ac{mcu} which is deployable to different application contexts: it is \emph{built for reuse}. Therefore the manufacturer first assumes the safety requirements that could be allocated from the system level and architecture around the component. These assumptions are necessary to develop the \ac{mcu} internal safety architecture. The system integrator has to follow the manufacturers assumptions and recommendations to preserve the integrity of the \ac{mcu} safety architecture in the final system context. For ISO 26262 compliant \acp{mcu} typically an additional document type is released in order to inform the integrator about the ISO 26262 related information essential for system integration activities: the \emph{safety manual} or \emph{safety application note}. In ISO 26262-10:2012 section A.3.10 an example on the content of the safety manual is given. 



As only suggestions for the safety manual content is provided, it still worth to examine which aspects have been realized in published documents. In order to assess the potential benefits of the safety manual in an avionic \ac{cots} component assurance process, the content of a representative probe of three manuals from three different vendors was analysed. The selected \acp{mcu} target ISO 26262 \ac{asil} D. They have been selected to increase the value of the scope of this article and not if they are really suitable for the avionics industry. Thus, no analysis has been performed to check the suitability of these devices for avionics due to e.g. cosmic radiation or other environmental or functional issues such as \emph{correct} set of interfaces. The selected \acp{mcu} with respective safety manuals are: NXP MPC5744P~\cite{NXP.062014}, ST SPC56ELx~\cite{ST.012018} and TI TMS570LC4x~\cite{TI.2016}. The content analysis of these manuals resulted in the following two major topics of interest that can be found in each of the examined safety manual in different level of detail:

\begin{itemize}
	\item \ac{mcu} safety architecture: It describes how random hardware fault management is separated between internal hardware diagnostics and additional software diagnostics. The examined \acp{mcu} employ a three layered approach:
	\begin{enumerate}
		\item All hardware blocks required for software execution are equipped with the highest degree of diagnostic coverage by hardware safety mechanisms. Two cores operate in delayed lock step and data transfers between memory and the processing cores are protected by end-to-end \ac{ecc}. This shall assure, that the software execution is not impacted by random hardware faults. 
		\item Based on the integrity of software execution, peripheral functions are mainly assured by software safety mechanism e.g. informational redundancy on application layer protocols. 
		\item Debug functions should not be used in an operational safety-related system, thus no diagnostics are provided and recommended respectively. 
	\end{enumerate}
	Worst case fault recognition times of hardware diagnostics are stated together with the failure indication and handling by entering safe states of the \ac{mcu}. 
	\item Hardware and software requirements on system level: Here the assumptions are explained which have to be followed by the system integrator. Hardware requirements define the functionality of external hardware safety mechanism like supervision of the power supply. Software requirements describe the correct way to utilize the internal hardware safety mechanisms and how software could improve the diagnostic coverage depending on the used \ac{mcu} hardware functions in the safety-related system. 
	
\end{itemize}

The avionic manufacturer could benefit from the same aspects as the automotive system integrator: At first from the ISO 26262 certified development process of the manufacturer and the process-requirements documented in the ISO 26262 respectively. At second, the additional information from the safety manual may be used. It can be assumed that the \ac{aeh} supplier may get further support from the \ac{mcu} manufacturer only in a limited scope, if necessary. However, these are the only public available information that can be additionally reused in particular for ISO 26262 compliant \acp{mcu} in the \ac{cots} assurance evaluation process described in the next section.


\subsection{\ac{cots} Component Assurance of ISO 26262 compliant \acp{mcu}}
In section \ref{sec:cotsAssurance}, \ac{cots} assurance activities were outlined on the basis of recommendations from~\cite{EASA.9.5.20121} for simple and complex/highly complex \acp{mcu}. The presented selection of ISO 26262 compliant \acp{mcu} in section \ref{sec:isoSpec} cannot be classified as simple\footnote{It is assumed that the full functional scope of the \ac{mcu} is used and in that case it will be not practical to verify it on that extent on device level.} and \acp{mcu} aiming at an even lower \ac{asil} level like \ac{asil} A or B are often based on more complex architectures. For that reason and to examine all benefits from the ISO compliance statement for every assurance activity, a classification of highly complex is assumed. The \ac{cots} component assurance activities have to be conducted by the \ac{aeh} supplier and some of them are achievable with minimal or no additional support by the \ac{mcu} manufacturer. These activities have to be excluded from the evaluation because they can be accomplished with \acp{mcu} in general and to claim these as ISO 26262 specific benefits would falsify the assessment results. Thus the following activities were omitted from the evaluation:

\begin{itemize}
	\item (1) Describing the \ac{cots} component characteristics in order to classify the \ac{mcu} as simple/complex/highly complex is feasible on basis of the usual public available hardware documentation.
	\item (2) Archiving of collected device data like errata notes or user manuals can be performed without help of the \ac{mcu} manufacturer
	\item (5) For usage the domain validation (part of activity (5)), the avionic system developer is responsible. Validation means, that a determined usage domain has to be checked if they do not contradict any higher level requirements from system/hardware/software. It is like requirements validation, to check if a low level requirement is a valid refinement of a higher level requirement. The \ac{cots} component manufacturer is not required for that task.
	\item (8) Documentation of past experience made with the \ac{mcu} during the \ac{aeh} development shall substantiate the robustness and maturity in the field. The \ac{mcu} manufacturers are not involved in this action.
	\item (15) Architectural mitigation techniques addressing common modes on device level. They are implemented during system development and are on a higher level than the \ac{mcu} itself\footnote{Note that on-chip \ac{mcu} architectural mitigation techniques cannot be credited for common mode issues.}. No additional support for this work can be requested from \ac{cots} component manufacturers.
\end{itemize}

Table \ref{tab:results} gives an overview of the evaluation results. The considered assurance activities can make use of additional \ac{mcu} artefacts in particular. They are assigned according to the identified top level arguments of section \ref{sec:cotsAssurance} and arranged in two groups resp.: \emph{Yes} if a \ac{cots} component assurance activity benefits from the ISO 26262 compliance statement and \emph{no} if that is not the case. 

\begin{table}[]
	\caption{Evaluation Results Overview}
	\label{tab:results}
	\centering
	\scriptsize
	\begin{tabularx}{1\linewidth}{m{3.6cm} Xc}
		\toprule
		\bfseries Top Level Argument & \bfseries Assurance Activity & \bfseries Benefits by ISO 26262  \\\midrule
		\multirow{6}{3.6cm}{1. The component performs as described by the manufacturer without anomalous behaviour.}& (3): Quality management and production&\multirow{2}{*}{No}\\
		&(13),(14): \ac{pse}&\\\cmidrule(rl){3-3}\cmidrule(rl){2-2}
		&(4), (5), (16)\footnotemark: Usage domain&\multirow{3}{*}{Yes}\\
		&(6), (7): Errata management&\\
		&(9), (10): Configuration management&\\\midrule
		2. The component as used satisfies the \ac{aeh} \newline requirements.&(11), (12): Integration&Yes\\
		\bottomrule
	\end{tabularx}
\end{table}
\footnotetext{Partitioning considerations were allocated to the usage domain analysis.}

For argument 1 no benefits can be directly asserted for activity (3), (13) and (14). Quality management and production process requirements in (3) can not be claimed to be defined by the ISO 26262. However, in a comprehensive ISO 26262 assessment process by a third party these aspects should also be checked.  Activity (13) and (14) require the documentation of the \ac{pse}. The ISO 26262 also introduces a \emph{proven in use argument} to claim a sufficient safety integrity, but no activities are defined that the \ac{mcu} manufacturer has to document the usage of their products in the automotive field. It is notable, that \ac{mcu} usage in the automotive safety critical sector is creditable if it can be adequately demonstrated. 

The determination (4) and verification (5) of the usage domain profits from detailed data descriptions in the safety manuals including disabling on chip functions, test of activated functions, implementation hints, mandatory requirements, assumptions, and initial configurations. Safety mechanisms described in the safety manual can also be utilized in usage domain verification tasks. Taking into account errata documents during system integration is demanded in the examined safety manual~\cite{NXP.062014,ST.012018,TI.2016}. They are published and sufficiently prepared in order to allow the system integrator to determine possible safety implications. Therefore, the errata management activities (6) and (7) should have an advantage by using a ISO 26262 compliant \ac{mcu}. Assurance activities (9) and (10) request an adequate configuration management or change description approach by the \ac{mcu} manufacturer and additional change impact analysis by the \ac{aeh} developer. According to ISO 26262 part 8 a configuration management and change management plan shall be provided by the \ac{mcu} manufacturer. In the safety manuals or errata documents the applicable device revision or product configurations are clearly stated. It is therefore assumed that \ac{cots} manufacturer’s configuration management is available and in good shape.

For argument 2 table \ref{tab:results} shows less assurance which activities benefit from an ISO 26262 compliant \ac{mcu}. Actually, most \ac{aeh} requirements are already determined and verified on device level in activity (4) and (5). Usage domain determination is a mapping of \ac{aeh} requirements on basis of the adequate configuration and usage of the \ac{mcu}. So the actual function and properties on device level designed by the manufacturer are reused as \ac{aeh} requirements. In activity (11) \ac{aeh} requirements from a higher level like \ac{lru} or \ac{cba} level allocated to the component have to be verified and validated. The device level description in the safety manual for I/O functions and software requirements may help in the validation and verification process for correct integration of the device. The assurance activity (12) demands a clear understanding of possible device failure modes and rates depending on its configuration. The safety manuals will help in this activity. Several failure scenarios are covered in these documents and failure rate calculations are one of the main topic of ISO 26262 hardware development.


\section{Conclusion}\label{sec:conlusion}
In this article an insight was given in the differences of assurance approaches for \ac{aeh} especially when equipped with complex \ac{cots} \acp{mcu}. Based on~\cite{EASA.9.5.20121} a new structured overview was presented for the \ac{cots} hardware component assurance activities. Currently, no industry consensus standard or recommendation from certification authorities is available that brings all necessary \ac{cots} assurance aspects together in an integrated approach ~\cite{AFE75ProjectManagementCommittee.}. Therefore the presented assurance activities are supposedly not complete. However, the selected assurance activities provide an adequate foundation for the evaluation of possible benefits of ISO 26262 compliant \acp{mcu} during the assurance process. Specifics of ISO 26262 compliant \acp{mcu} were described to identify the aspects that could be reused. The evaluation concentrates on assurance activities where additional support by the \ac{mcu} manufacturer is most helpful. It could be demonstrated that an ISO 26262 compliant \ac{mcu} is beneficial for the \ac{aeh} manufacturer by conducting certain assurance activities. However, the magnitude of these advantages depend on the dedicated context in which the \ac{mcu} should be integrated.


\section*{Acknowledgment}
This paper is sponsored by the Airbus Defense and Space endowed professorship "System Technology for safety-related Applications" supported by "Stifterverband f\"{u}r die Deutsche Wissenschaft e.V.".

MDH’s work in this paper is supported by the Swedish Knowledge Foundation within the project DPAC.
\bibliographystyle{splncs}
\bibliography{Literatur}
\acuse{oem}

\end{document}